\documentstyle[12pt,epsfig]{article}
\begin{document}

\newcommand{\rp}{\right)}
\newcommand{\lp}{\left(}
\newcommand \be  {\begin{equation}}
\newcommand \bea {\begin{eqnarray}}
\newcommand \ee  {\end{equation}}
\newcommand \eea {\end{eqnarray}}

\title{Evaluation of the quantitative prediction of a trend
reversal on the Japanese stock market in 1999.}

\author{Anders Johansen$^1$ and Didier Sornette$^{1,2,3}$ \\
$^1$ Institute of Geophysics and
Planetary Physics\\ University of California, Los Angeles, California 90095\\
$^2$ Department of Earth and Space Science\\
University of California, Los Angeles, California 90095\\
$^3$ Laboratoire de Physique de la Mati\`{e}re Condens\'{e}e\\ CNRS UMR6622 and
Universit\'{e} de Nice-Sophia Antipolis\\ B.P. 71, Parc
Valrose, 06108 Nice Cedex 2, France}


\maketitle

\begin{abstract}

\noindent In January 1999, the authors published a quantitative prediction that
the Nikkei index should recover from its 14 year low in January 1999 and reach
$\approx 20500$ a year later. The purpose of the present paper is to evaluate
the performance of this specific prediction as well as the underlying model:
the forecast, performed at a time when the Nikkei was at its lowest (as we
can now judge in hindsight), has correctly captured the change of trend as
well as the quantitative evolution of the Nikkei index since its inception.
As the change of trend from sluggish to recovery
was estimated quite unlikely by many observers at that time, a Bayesian
analysis shows that a skeptical (resp. neutral) Bayesian
sees her prior belief in our model amplified into a posterior belief $19$ times
larger (resp. reach the $95\%$ level).
\vspace{1cm}

\noindent keywords: Stock market; Log-periodic oscillations; scale invariance;
prediction; Gold; Nikkei; Herding behaviour.
\end{abstract}
\newpage

\pagenumbering{arabic}

Following the general guidelines proposed in \cite{manisfesto}, the authors
made in January 1999 public through the Los Alamos preprint server
\cite{lanlsub} a quantitative prediction stating that the Nikkei index should
recover from its 14 year low (actually 13232.74 on 5 Jan 1999) and reach
$\approx 20500$ a year later corresponding to an increase in the index of
$\approx 50\%$. Furthermore, this prediction was mentioned in a
wide-circulation journal which appeared in May 1999 \cite{DSStaupredNi}.

\begin{figure}
\begin{center}
\epsfig{file=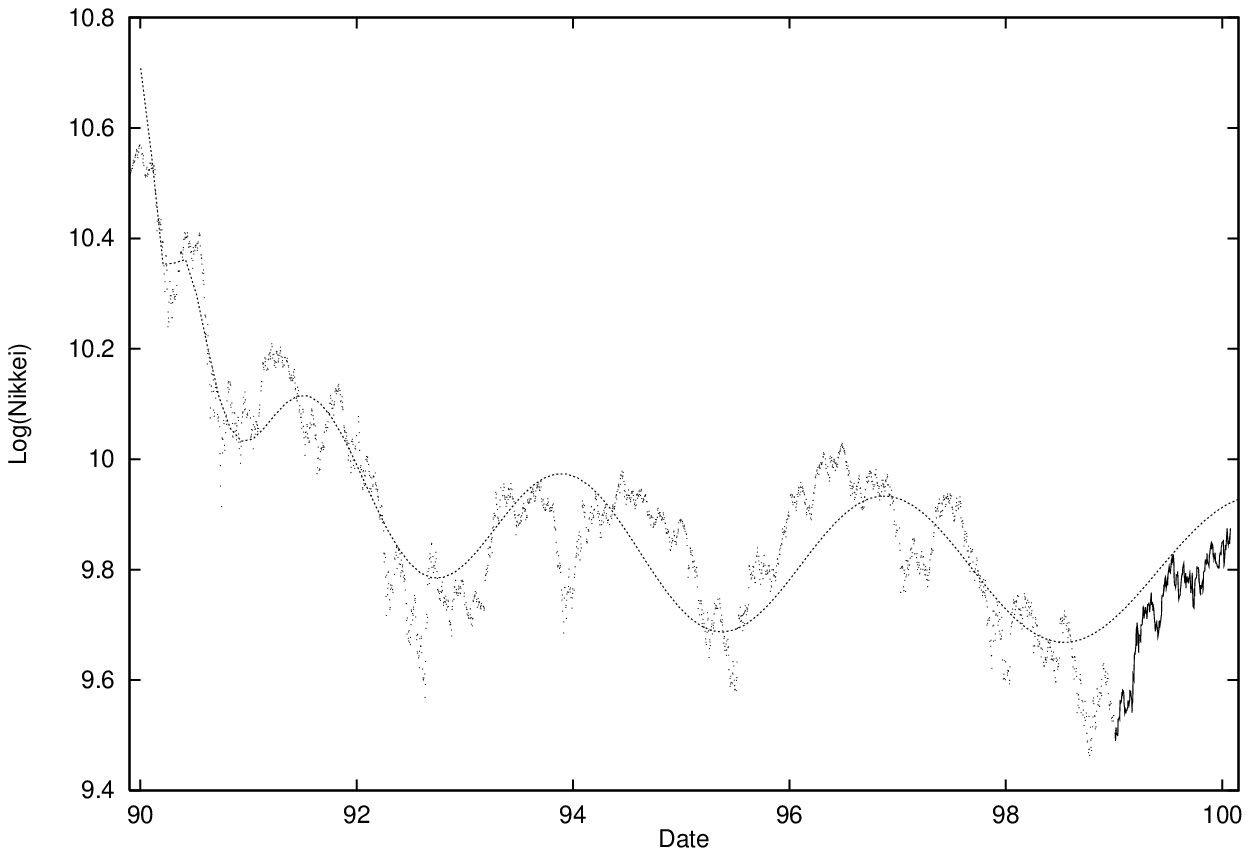,width=12cm}
\caption{\protect\label{predfig}. Logarithm of the Nikkei Index compared to
equation \protect\ref{3feq}. The dots are the data used in the fit of equation
(\ref{3feq}) being the ticked line and covers the 9 year period from 31 Dec.
1989 to 31 Dec. 1998. The solid line is the actual behaviour of the Nikkei
{\em after} the last point used in the fit and covers the period 1 Jan. 1999
to 28 Jan. 2000. The prediction was made public on the 25 Jan. 1999
\protect\cite{lanlsub}. See \protect\cite{lanlsub} for details of the fit.}
\vspace{5mm}
\epsfig{file=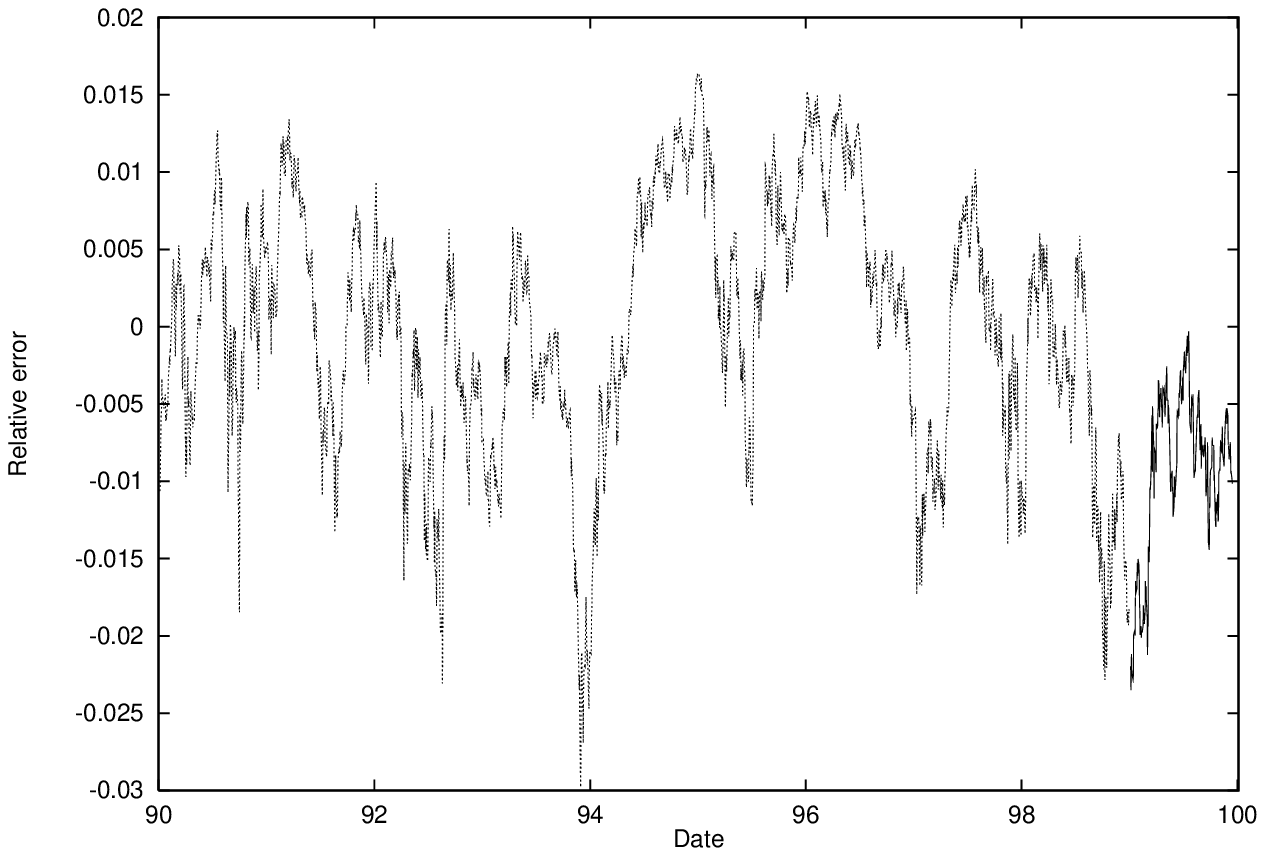,width=12cm}
\caption{\protect\label{relerrorfig} The relative error between the fit with
equation (\protect\ref{3feq}) and the data. The ticked line if the relative
error between the fit and the data used in the fit. The solid line is the error
between the prediction and the actual data.}
\end{center}
\end{figure}

Specifically, based on a third-order ``Landau'' expansion
\be
\frac{d F\lp \tau \rp}{d\log \tau}=\alpha F\lp \tau \rp +\beta
|F\lp \tau \rp|^2 F\lp \tau \rp  + \gamma |F\lp \tau \rp|^4 F\lp \tau
\rp\ldots
\label{3expan}
\ee
in terms of $\tau \equiv t-tc$, where $t_c =$ 31 Dec. 1989 is the time of
the all-time
high of the Nikkei index,
the authors arrived at the equation
$$
\log\lp p(t)\rp \approx A' + \frac{\tau^\alpha}{\sqrt{1+\left(\frac{\tau}
{\Delta_t}\right)^{2\alpha} + \left(\frac{\tau}{\Delta_t'}\right)^{4\alpha}}}
$$
\be \label{3feq}
\left\{B'+ C'\cos\left[\omega\log \tau +
\frac{\Delta_\omega}{2\alpha}\log\left(1+\left(\frac{\tau}
{\Delta_t}\right)^{2\alpha}\right)+
\frac{\Delta_\omega'}{4\alpha}\log\left(1+\left(\frac{\tau}
{\Delta_t'}\right)^{4\alpha}\right) + \phi\right]\right\}~,
\ee
describing the time-evolution of the Nikkei Index $p(t)$. Equation (\ref{3feq})
was then fitted to the Nikkei index in the time interval from the beginning of
1990 to the end of 1998, {\it i.e.}, a total of 9 years. Extending the curve
beyond 1998 thus provided us with a quantitative prediction for the future
evolution of the Index. In figure \ref{predfig}, we compare the actual and
predicted evolution of the Nikkei over 1999. We see that not only did the
Nikkei experience a trend reversal as predicted, but it has also followed
the quantitative prediction with rather impressive precision, see figure
\ref{relerrorfig}. It is important to note that the error between the curve
and the data has not grown after the last point used in the fit. This tells
us that the prediction has performed well so far. Furthermore, since the
relative error between the fit and the data is within $\pm 2\%$ over a
time period of 10 years, not only has the prediction performed well, but also
the underlying model. This analysis represents the correct quantitative
evaluation of the performance of the model as well as its predictive power
on the Nikkei Index over a quite impressive time-span of 10 years.

We wish to stress that the fulfilling of our prediction is even more
remarkable than the comparison between the curve and the data indicates.
This, since it included {\it a change of trend}: at the time when the
prediction was issued, the market was declining and showed no tendency to
increase. Many economists were at that time very pessimistic and could not
envision when Japan and its market would rebounce. For instance, the well-known
economist P. Krugman \cite{Krugman} wrote July 14, 1998 in the Shizuoka Shimbun
at the time of the banking scandal ``the central problem with Japan right now
is that there just is not enough demand to go around - that consumers and
corporations are saving too much and borrowing too little... So seizing these
banks and putting them under more responsible management is, if anything,
going to further reduce spending; it certainly will not in and of itself
stimulate the economy... But at best this will get the economy back to where
it was a year or two ago - that is, depressed, but not actually plunging.''
Then, in the Financial Times, January, 20th, 1999, P. Krugman wrote in an
article entitled ``Japan heads for the edge'' the following: ``...the story
is starting to look like a tragedy. A great economy, which does not deserve
or need to be in a slump at all, is heading for the edge of the cliff -- and
its drivers refuse to turn the wheel.'' In a poll of thirty economists
performed
by Reuters (the major news and finance data provider in the world)
in October 1998 \cite{poll}, only two economists predicted growth for
the fiscal year of 1998-99. For the year 1999-2000 the prediction was a
meager 0.1\% growth. This majority of economists said that
``a vicious cycle in the economy was unlikely to disappear any
time soon as they expected little help from the government's economic stimulus
measures... Economists blamed moribund domestic demand, falling prices,
weak capital spending
and problems in the bad-loan laden banking sector for dragging down the
economy.''

Nevertheless, we predicted a $\approx 50\%$ increase of the market in the
next 12 months assuming that the Nikkei would stay within the error-bars of
the fit. At the time of writing (3rd February 2000), the market is up by
$\approx 49.5\%$ and the error between the prediction and the curve has not
increased, see figure \ref{relerrorfig}. Predictions of trend reversals is
noteworthy difficult and unreliable, especially in the linear framework of
auto-regressive models used in standard economic analyses. The present
nonlinear framework is well-adapted to the forecasting of change of trends,
which constitutes by far the most difficult challenge posed to forecasters.
Here, we refer to our prediction of a trend reversal
within the strict confine of equation (\ref{3feq}): trends are limited
periods of times when the oscillatory behavior shown in figure \ref{predfig}
is monotonous. A change of trend thus corresponds to crossing a local maximum
or minimum of the oscillations.

We report one case. In the standard ``frequentist'' approach to probability
\cite{jeffreys}
and to the establishment of statistical confidence, this bears essentially
no weight and should be discarded as story telling. We are convinced that
the ``frequentist'' approach is unsuitable to assess the quality of such a
unique
experiment as presented here of the prediction of a global financial indicator
and that the correct framework is Bayesian. Within the Bayesian framework,
the probability
that the hypothesis is correct given the data can be estimated, whereas this
is excluded by construction in the standard ``frequentist''  formulation,
in which
one can only calculate
the probability that the null-hypothesis is wrong, not that the alternative
hypothesis is correct (see also \cite{Bayes} for recent introductory
discussions).

Bayes' theorem states that
\begin{equation}
P(H_i|D) = \frac{P(D|H_i) \times P(H_i)}{\sum_j P(D|H_j)P(D_j)}\,.
\label{eq:bayes2}
\end{equation}
where the sum in the denominator runs over all the different
conflicting hypothesis.  In words,
equation (\ref{eq:bayes2}) estimates that the probability, that hypothesis
$H_i$ is
correct given the data $D$, is proportional to the probability $P(D|H_i)$ of
the data given the hypothesis $H_i$ multiplied with the prior belief $P(H_i)$
in the hypothesis $H_i$ divided with the probability of the data. In the
present context, we use only the two hypotheses $H_1$ and $H_2$ that our
prediction of a trend reversal is correct or that it is wrong. For the data,
we take the change of trend from bearish to bullish. We now want to estimate
whether the fulfillment of our prediction was a ``lucky one''. We quantify
the general atmosphere of disbelief that Japan would recover by the value
$P(D|H_2) = 5\%$ for the probability that the Nikkei will change trend while
disbelieving our model. We assign the classical confidence level of $P(D|H_1)
= 95\%$ for the probability that the Nikkei will change trend while believing
our model.

Let us consider a skeptical Bayesian with prior probability (or belief)
$P(H_1)
= 10^{-n}$, $n\ge 1$ that our model is correct. From (\ref{eq:bayes2}), we get
\be
P(H_1|D) = {0.95 \times 10^{-n} \over 0.95 \cdot 10^{-n}  + 0.05 \times
(1-10^{-n})}~ .
\ee
For $n=1$, we see that her posterior belief in our model has been
amplified compared to her prior belief by a factor $\approx 7$ corresponding to
$P(H_1|D) \approx 70\%$. For $n=2$, the amplification factor is $\approx 16$
and hence $P(H_1|D) \approx 16\%$. For large $n$ (very skeptical Bayesian),
we see that her posterior belief in our model has been amplified compared
to her
prior belief by a factor $0.95/0.05 = 19$.
Alternatively, consider a neutral Bayesian
with prior belief $P(H_1) = 1/2$, {\it i.e.}, a priori she considers equally
likely that our model is correct or wrong. In this case, her prior belief
is changed
into the posterior belief equal to
\be
P(H_1|D) = {0.95 \cdot {1 \over 2} \over 0.95 \cdot {1 \over 2}  +
0.05 \cdot {1 \over 2}} = 95\%~.
\ee
This means that this single case is enough to convince the neutral Bayesian.

We stress that this specific application of Bayes' theorem only deals with a
small part of the model, {\it i.e.}, the trend reversal. It does not establish
the significance of the quantitative description of {\em 10 years} of data
(of which the last one was unknown at the time of the prediction) by
the proposed model within a relative error of $\approx \pm 2\%$.

A question that remains is how far into the future will the Japanese stock
market continue to follow equation (\ref{3feq})? Obviously, the Nikkei Index
must ``break away'' at some point in the future even if there are no changes
in the overall behaviour and the underlying model thus remains valid. The
reason is that the prediction was made using a third order expansion. This
means that, as the parameter $\tau = t-t_c$ in equation (\ref{3feq})
continues to
increase, this approximation becomes worse and worse and a fourth order term
should be included. Presently, we are not ready to present the derivation
of such an
equation. Furthermore, we expect the numerical difficulties involved in
fitting an even more complex equation than equation (\ref{3feq}) to be
considerable.

Last, we would like to bring to the attention of the reader that not only can
bearish markets occasionally be described by the framework underlying equation
(\ref{3feq}). In fact, bullish markets exhibits such changes of regimes
even more frequently, see \cite{JSL}.

\vspace{1cm}

{\bf Acknowledgement} The authors wish to thank D. Stauffer for his
encouragement  both with respect to the original work of \cite{lanlsub} as
well as the present re-evaluation.

\begin{thebibliography}{}

\bibitem{manisfesto} A. Johansen and D. Sornette,
{\it Modeling the Stock Market prior to large crashes}, Eur. Phys. J. B,
{\bf 9}, 167-174. Available on http://www.nbi.dk/\~~johansen/pub.html

\bibitem{lanlsub} The prediction was made public on the 25 Jan. 1999 by
posting a preprint on the Los Alamos server, see
http://xxx.lanl.gov/abs/cond-mat/9901268 . The preprint was later published
as A. Johansen and  D. Sornette,
{\it Financial ``anti-bubbles'': log-periodicity in Gold and Nikkei collapses},
Int. J. Mod. Phys. C. {\bf 10},  563-575 (1999).

\bibitem{DSStaupredNi}  D. Stauffer, Monte-Carlo-Simulation mikroskopischer
B\"{o}rsenmodelle, Physikalische Bl\"{a}tter 55 (1999) 49.

\bibitem{Krugman} The Official Paul Krugman Web Page:
http://web.mit.edu/krugman/www/

\bibitem{poll} Reported in Indian Express on the 15 Oct., see\\
http://www.indian-express.com/fe/daily/19981016/28955054.html

\bibitem{jeffreys} H. Jeffreys,
{\it Theory of Probability}, ~3rd ed. (Oxford University Press, 1961)).

\bibitem{Bayes} D. Malakoff,
{\it Bayes Offers a 'New' Way to Make Sense of Numbers},
Science {\bf 286}, 1460-1464 (1999);
{\it A Brief Guide to Bayes Theorem}, ibid {\bf 286}, 1461 (1999);
G. D'Agostini, G.,
{\it Teaching statistics in the physics curriculum: Unifying and clarifying
role of subjective probability}, Am. J. Phys. {\bf 67}, 1260-1268 (1999).

\bibitem{JSL} A. Johansen, D. Sornette and O. Ledoit,
{\it Predicting Financial Crashes Using Discrete Scale Invariance},
J. of Risk, Vol 1 No. 4, pp.5-32 (1999).
Available on http://www.nbi.dk/\~~johansen/pub.html;
A. Johansen and D. Sornette,
{\it Log-periodic power law bubbles in Latin-American and Asian markets
and correlated anti-bubbles in Western stock markets: An empirical study}.
Subm. to J. Empirical Finance. Preprint available on
http://www.nbi.dk/\~~johansen/pub.html



\end{thebibliography}
\end{document}